\documentclass[preprint,12pt]{elsarticle}

\usepackage{graphicx}
\usepackage{subfigure}

\usepackage{amssymb}

\journal{Solid State Communications}

\begin{document}

\begin{frontmatter}



\title{
 Computational Materials Design for Superconductivity
 in Hole-Doped Delafossite CuAlO$_2$: Transparent Superconductors
}

\author{Akitaka Nakanishi\corref{cor}}
\ead{nakanishi@aquarius.mp.es.osaka-u.ac.jp}
\author{Hiroshi Katayama-Yoshida}
\cortext[cor]{Tel.:+81-6-6850-6504. Fax:+81-6-6850-6407.}
\address{Graduate School of Engineering Science,
 Osaka University, Toyonaka, Osaka 560-8531}

\begin{abstract}
We have calculated the superconducting critical temperature $T_{\rm c}$
 of hole-doped delafossite CuAlO$_2$ based on the first-principles calculations.
According our calculation, $0.2\sim0.3$ hole-doped CuAlO$_2$ 
 can become a phonon-mediated high-$T_{\rm c}$ superconductor with $T_{\rm c}\simeq50$\,K.
In the hole-doped CuAlO$_2$,
 the A$_1$L$_1$ phonon mode that stretches O-Cu-O dumbbell
 has a strong interaction with electrons of the flat band in
 Cu 3d$_{3z^2-r^2}$ and the O 2p$_{z}$ anti-bonding $\pi$-band.

\end{abstract}


\end{frontmatter}


\section{Introduction}
Kawazoe {\it et al.} have discovered the delafossite structure of
 CuAlO$_2$ is the transparent $p$-type conductor
 without any intentional doping.\cite{Kawazoe1997}
Transparent $p$-type conductors such as CuAlO$_2$ are
 rare and absolutely necessary for the $p$-$n$ junction of the transparent conductors
 and high-efficient photovoltaic solar cells.
Many applications of CuAlO$_2$
 for flat panel displays, photovoltaic solar-cells, touch panels, and
 high efficiency thermoelectric-power materials
 with about 1\% hole-doping\cite{Funashima2004,Yoshida2009} are expected.

Recently,
 Katayama-Yoshida {\it et al.} have suggested
 a new application of CuAlO$_2$
 for transparent superconductivity
 and high-efficient thermoelectric-power material 
 with a large Seebeck coefficient caused
 by the flat band.\cite{Yoshida2003}
They have simulated
 the $p$-type doped CuAlO$_2$
 by shifting the Fermi level rigidly
 with FLAPW method,
 and proposed that
 the nesting Fermi surface may cause
 a strong electron-phonon interaction
 and a transparent superconductivity for visible light
 due to the large band gap ($\sim3.0$eV).
But,
 the calculation of superconducting critical temperature $T_{\rm c}$
 is not carried out.
In this study,
 we calculated the electron-phonon interaction and
 the $T_{\rm c}$ of $p$-type doped CuAlO$_2$
 based on the first principles calculation
 with the density functional perturbation theory.\cite{Baroni2001}
We found that the $T_{\rm c}$ goes up to about 50\,K
 due to the strong electron-phonon interaction
 and high phonon frequency caused 
 by the two dimensional flat band
 in the top of the valence band.

\section{Calculation Methods}
The calculations are performed
 within the density functional theory\cite{Hohenberg1964,Kohn1965}
 with a plane-wave pseudopotential method,
 as implemented in the Quantum-ESPRESSO code.\cite{Giannozzi2009}
We employed the Perdew-Burke-Ernzerhof
 generalized gradient approximation (GGA)
 exchange-correlation functional\cite{Perdew1996}
 and ultra-soft pseudopotentials.\cite{Vanderbilt1990}
For the pseudopotentials,
 Cu 3d electrons were also included
 in the valence electrons.
In reciprocal lattice space integral calculation,
 we used $8\times8\times8$ (electron and phonon)
 and $32\times32\times32$ (average at Fermi level)
  ${\bf k}$-point grids
 in the Monkhorst-Pack grid.\cite{Monkhorst1976}
The energy cut-off for wave function is 40  Ry and
 that for charge density is 320  Ry.
These ${\bf k}$-point meshes are fine
 enough to achieve convergence
 within 0.1 mRy/atom in the total energy.
The $32\times32\times32$ mesh
 for average at Fermi level
 is enough to achieve convergence
 in the electron-phonon interaction
 and the superconducting critical temperature.
The differences between
 results of $32\times32\times32$ k-points mesh
 and those of $64\times64\times64$ one
 are less than 1\%.

\begin{figure}[htbp]
  \begin{center}
    \includegraphics{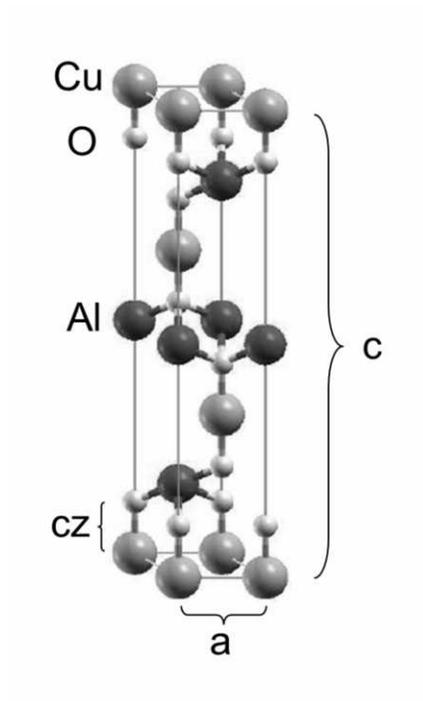}
    \caption{The crystal structure of delafossite CuAlO$_2$.}
    \label{fig:str}
  \end{center}
\end{figure}

The delafossite structure belongs to the space group R$\bar3$m (No.166)
 and is represented by cell parameters $a$ and $c$,
 and internal parameter $z$ (See Fig. \ref{fig:str}).
These cell parameters and internal parameter were optimized by 
 the constant-pressure variable-cell relaxation using
 the Parrinello-Rahman method\cite{Parrinello1980}
 without any symmetry requirements.
The results of relaxation ($a=2.861$\AA,$c/a=5.969$ and $z=0.1101$)
 agree very well with the experimental data
 ($a=2.858$\AA,$c/a=5.934$ and $z=0.1099$ \cite{Nie2002,Buljan1999}).

In this study,
 some properties of hole-doped CuAlO$_2$ are approximated
 because it is difficult for first-principles calculation
 to deal with the doped system exactly.
Let's take the electron-phonon interaction $\lambda$ for example.
$\lambda$ is defined as follows:

\begin{equation}
 \lambda
 =
 \sum_{\nu{\bf q}}
 \frac{
 2N(\varepsilon_{\rm F})
 \sum_{{\bf k}}|M_{{\bf k,k+q}}^{\nu{\bf q}}|^2
 \delta(\varepsilon_{\bf k}-\varepsilon_{\rm F})
 \delta(\varepsilon_{{\bf k+q}}-\varepsilon_{\rm F})
 }
 {
 \omega_{\nu{\bf q}}\sum_{{\bf kq'}}
 \delta(\varepsilon_{\bf k}-\varepsilon_{\rm F})
 \delta(\varepsilon_{{\bf k+q'}}-\varepsilon_{\rm F})
 }.
\label{eqn:lambda}
\end{equation}
(1)
 For the non-doped CuAlO$_2$,
 we calculated the dynamical matrix,
 the phonon frequency $\omega_{\nu{\bf q}}$ 
 and the electron-phonon matrix $M_{{\bf k,k+q}}^{\nu{\bf q}}$.
(2)
 For the doped CuAlO$_2$,
 we calculated the Fermi level $\varepsilon_{\rm F}$ and
 the density of states at the Fermi level $N(\varepsilon_{\rm F})$
 with the number of valence electrons reduced
 using the eigenvalues $\varepsilon_{\bf k}$ of the non-doped system.
(3)
 By using the results of (1) and (2),
 we calculated the electron-phonon interaction $\lambda$
 and the other superconducting properties.
This approximation is based on the rigid band model
 and the idea that the doping does not greatly change
 the phonon band structures.
In this study,
 we show the results of $0.1\sim1.0$ hole-doped CuAlO$_2$.

\section{Calculation Results and Discussion}
\begin{figure}[htbp]
\begin{center}
  \includegraphics{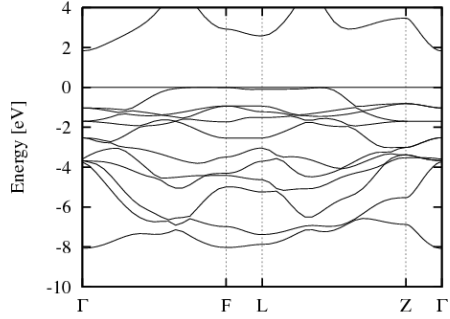}
  \caption{Band structure of CuAlO$_2$.}
\label{fig:band_CuAlO2}
\end{center}
\end{figure}

\begin{figure}[htbp]
\begin{center}
  \includegraphics{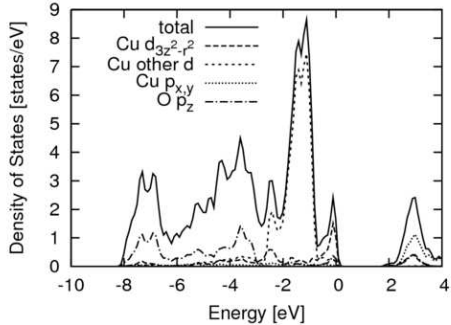}
  \caption{
  Total density of states(DOS) and projected DOS.
  }
\label{fig:pdos_CuAlO2}
\end{center}
\end{figure}

Before the discussion of the superconducting critical temperature,
 let us see the electronic structure.
Fig. \ref{fig:band_CuAlO2} and
 Fig. \ref{fig:pdos_CuAlO2} show
 the electronic band structure and
 density of states (DOS) of non-doped CuAlO$_2$.
The top of the valence band of CuAlO$_2$
 is flat due to the two dimensionality
 in O-Cu-O dumbbell array,
 and has a small peak 
 in the DOS of the valence band.
This peak is mainly constructed
 by the two-dimensional $\pi$-band of 
 Cu 3$d_{3z^2-r^2}$-O 2$p_z$ anti-bonding state.

\begin{figure}[htbp]
\begin{center}
  \includegraphics{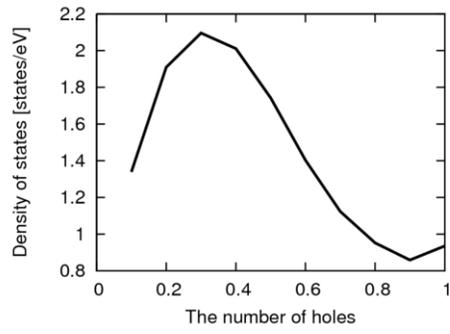}
  \caption{
  The number of holes vs density of states at the Fermi level.
  }
\label{fig:dos_CuAlO2}
\end{center}
\end{figure}

Fig. \ref{fig:dos_CuAlO2} shows
 the DOS at the Fermi level calculated
 with the number of valence electron reduced.
According to this figure and Fig. \ref{fig:pdos_CuAlO2},
 the number of holes $N_{\rm h}=0.3$ corresponds to that Fermi level
 which is located at the top of the peak in the DOS,
 and $N_{\rm h}=0.9$ corresponds to
 that Fermi level which is located
 at the bottom of the DOS. 

We calculated the superconducting critical temperature
 by using the Allen-Dynes modified McMillan's formula.\cite{McMillan1968,Allen1975} 
According to this formula,
 $T_{\rm c}$ is given by three parameters:
 the electron-phonon interaction $\lambda$, 
 the logarithmic averaged phonon frequency $\omega_{\log}$, 
 and the screened Coulomb interaction $\mu^{\ast}$, in the following form.
\begin{eqnarray}
  T_{\rm c}&=&\frac{\omega_{\log}}{1.2}
  \exp \left( -\frac{1.04(1+\lambda )}
  {\lambda-\mu^{\ast}(1+0.62\lambda )} \right). \\
  \omega_{\log} 
  &=&\exp \left(\frac{2}{\lambda}\int_0^\infty
  d\omega\frac{\alpha^2F(\omega)}{\omega}\log\omega\right),
\end{eqnarray}
Here, $\alpha^2F(\omega)$ is the Eliashberg function.
$\lambda$ and $\omega_{\rm log}$ are obtained by 
 the first-principle calculations using
 the density functional perturbation theory.
As for $\mu^{\ast}$,
 we assume the value $\mu^{\ast}=0.1$.
This value holds for weakly correlated materials
 due to the electronic structure
 of lightly hole-doped Cu$^+$ (d$^{10}$).

\begin{figure}[htbp]
\begin{center}
  \includegraphics{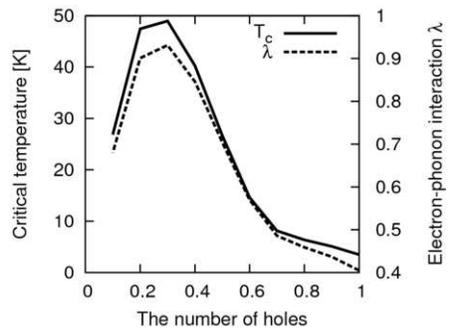}
  \caption{
  Superconducting critical temperature and
   electron-phonon interaction $\lambda$.
  }
\label{fig:tc_CuAlO2}
\end{center}
\end{figure}

The calculated result of $T_{\rm c}$ and $\lambda$
 as a function of $N_{\rm h}$
 is shown in Fig. \ref{fig:tc_CuAlO2}.
In our calculation, the lightly doped CuAlO$_2$
 ($N_{\rm h}=0.2\sim0.3$) has $T_{\rm c}\simeq 50$\,K.
This $T_{\rm c}$ is the highest
 among phonon-mediated superconductors.
In addition,
 the $T_{\rm c}$ can be increased
 by other purely attractive electron-electron interaction mechanisms:
 for example,
 charge-excitation-induced\cite{Yoshida2008}
 or exchange-correlation-induced\cite{Yoshida1985}
 negative effective U system
 in the Cu$^+$ (d$^{10}$) electronic structure
 with lightly hole-doping.
The heavily doped CuAlO$_2$ ($N_{\rm h}=0.6\sim1.0$)
 has $T_{\rm c}\simeq 10$\,K
 by reducing the electron-phonon interaction.

Let us examine the origin of the high $T_{\rm c}$
 of lightly hole-doped CuAlO$_2$.
In this study,
 the critical temperature is determined
 by $\lambda$ and $\omega_{\log}$
 as mentioned above.
Table \ref{tab:tc_CuAlO2} shows $\lambda$ and $\omega_{\log}$.
The electron-phonon interaction $\lambda$
 at $N_{\rm h}=0.3$ is about 130\% larger
 than that at $N_{\rm h}=1.0$,
 while $\omega_{\log}$ at $N_{\rm h}=0.3$ is about 1\% larger
 than that at $N_{\rm h}=1.0$.
In addition,
 $\lambda$ affects $T_{\rm c}$ exponentially,
 while $\omega_{\log}$ affects $T_{\rm c}$ linearly.
Therefore,
 the high $T_{\rm c}$ is attributed to the strong electron-phonon interaction.

\begin{table} [tpd]
\begin{center}
\begin{tabular}{ccc}
  \hline \hline
  $N_{\rm h}$ & $\lambda$ & $\omega_{\log}[K]$ \\ \hline 
          0.3\hspace{0.3cm} &     0.931 &               789  \\
          1.0\hspace{0.3cm} &     0.405 &               778  \\ 
  \hline \hline
  \end{tabular}
  \label{tab:tc_CuAlO2}
  \caption{
  Electron-phonon interaction $\lambda$
  and logarithmic averaged phonon frequencies $\omega_{\log}$.
  $T_{\rm c}$ has max. and min. at $N_{\rm h}=0.3, 1.0$.
  }
\end{center}
\end{table}

\begin{figure}[tbp]
  \begin{center}
  \subfigure[]{
  \includegraphics{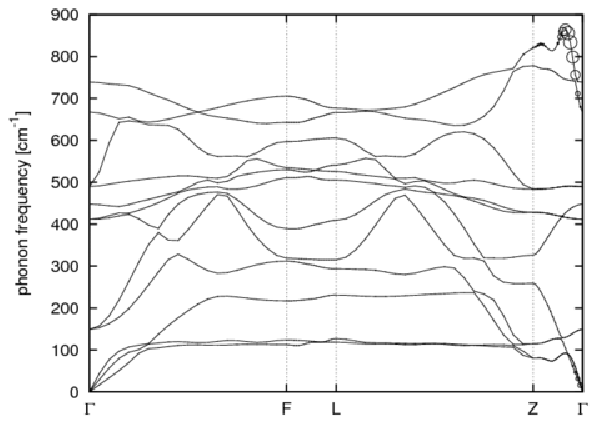}
  }
  \subfigure[]{
  \includegraphics{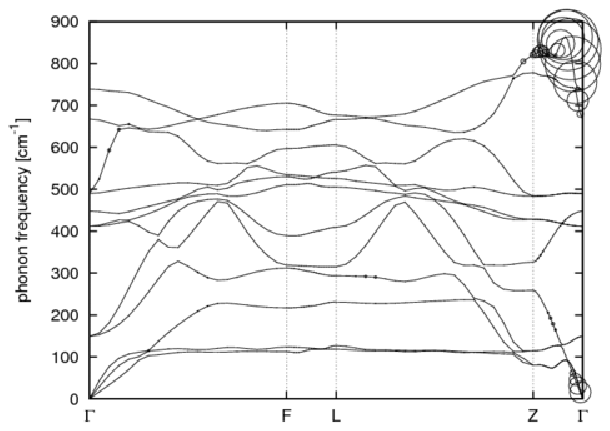}
  }
  \end{center}
  \caption{
  Phonon dispersions and electron-phonon interactions of hole-doped CuAlO$_2$.
  The radius of circle represents
   the strength of partial electron-phonon interaction $\lambda_{\nu{\bf q}}$.
  Note that many $\lambda_{\nu{\bf q}}$ are very small
   and their circles are no longer invisible.
  (a) The number of holes $N_{\rm h}=1.0$.
  (b) $N_{\rm h}=0.3$.
  }
  \label{fig:lambda_CuAlO2}
\end{figure}

In Fig.\ \ref{fig:lambda_CuAlO2},
 we show the phonon dispersion which
 show the strong two dimensionality
 with flat phonon dispersion.
In order to find that
 which phonon mode has a large contribution to the high $T_{\rm c}$,
 we introduce a partial electron-phonon interaction $\lambda_{\nu{\bf q}}$:
 the interaction of a phonon whose frequency is $\omega_{\nu{\bf q}}$.
Then, 
 $\lambda = \sum_{\nu {\bf q}} \lambda_{\nu{\bf q}}$.
In Fig.\ \ref{fig:lambda_CuAlO2},
 the $\lambda_{\nu {\bf q}}$ is shown
 by the radius of a circle on each phonon dispersion.
Since most of $\lambda_{\nu{\bf q}}$ are very small,
 their circles are no longer invisible
 in the figure \ref{fig:lambda_CuAlO2}.
This figure indicates that
 the highest mode on the $Z-\Gamma$ line
 has a large electron-phonon interaction.
In the case of $N_{\rm h}=0.3$,
 the sum of $\lambda_{\nu {\bf q}}$
 of the highest frequency mode is 0.407.
This value is about 44\% of
 total electron-phonon interaction $\lambda=0.931$.
The effective mode is the A$_1$L$_1$ phonon mode.
In this mode,
 the O atoms oscillate in the anti-phase
 within an O-Cu-O dumbbell.

As mentioned above,
 the CuAlO$_2$ has the Cu 3$d_{3z^2-r^2}$
 and the O 2$p_z$ electrons
 at the top of the valence band.
When $N_{\rm h}=0.2\sim0.3$,
 the electrons which make
 the O-Cu-O anti-bonding band are located
 at the Fermi level.
They have a strong interaction
 with the A$_1$L$_1$ phonon mode
 because their bonding direction is parallel to
 the oscillation direction of the A$_1$L$_1$ phonon mode.
Though the strong electron-phonon interaction,
 the O-Cu-O bonding of the delafossite structure
 is stable even under high pressure.\cite{Nakanishi2011a}
There is a strong possibility
 that the doped CuAlO$_2$ is stable and a superconductor.
When CuAlO$_2$ is heavily hole-doped,
 the total electron-phonon interaction decreases
 because the number of electrons which
 have a strong interaction decreases.

The top of valence band is constructed by
 the Cu 3$d_{3z^2-r^2}$ and
 the O 2$p_z$ anti-bonding $\pi$-band.\cite{Nakanishi2011a,Nakanishi2011b,Hamada2006}
The hole-doping makes the O-Cu-O coupling more strong.
Therefore,
 when the density of holes increases from 0.2 to 0.3,
 $\lambda$ and $\omega_{\rm log}$ does not change much
 ($\lambda=0.901\rightarrow0.931,\omega_{\rm log}=808\rightarrow789$K).

D. Huang and Y. Pan have investigated the intrinsic defects in CuAlO$_2$.\cite{Huang2010}
According to their study,
 vacancies at the Cu sites and substitutional Cu at the Al site are 
 most likely responsible for the $p$-type conductivity,
 and the transition levels of these defects are deep.
The rigid-band doping is possibly not realized in CuAlO$_2$.
Many theoretical proposals for superconductivity-in-semiconductors
(for example, LiBC\cite{Rosner2002}) are based on rigid band or other oversimplified models
 and not successful in experiment
 because doping concentrations cannot be realized or large doping levels cause material distortions
 not accounted for by these models.
Therefore,
 the calculated $T_{\rm c}\simeq50\,K$ may not be realized in experiment.
However,
 we believe our calculation results suggest the superconductivity potential of hole-doped CuAlO$_2$.

\section{Conclusions}
In summary,
 we calculated the superconducting critical temperature of
 the hole-doped delafossite CuAlO$_2$
 by shifting the Fermi level rigidly
 based on the first principles calculation.
The lightly hole-doped CuAlO$_2$ has
 Cu 3$d_{3z^2-r^2}$ and O 2$p_{z}$ anti-bonding $\pi$-band
 as the top of the valence band.
The electrons of this band have
 a strong electron-phonon interaction
 with the A$_1$L$_1$ phonon mode
 because the direction of O-Cu-O dumbbell is parallel to 
 the oscillation direction of the A$_1$L$_1$ phonon mode.
These findings suggest that hole-doped CuAlO$_2$ may be a superconductor.
We hope that our computational materials design of superconductivity
 will be verified by experiments very soon.
We can easily extend the present computational materials design
 to other delafossite structures of lightly hole-doped
 AgAlO$_2$ and AuAlO$_2$ which may have higher $T_{\rm c}$
 due to the strong electron-phonon interaction
 combined with the charge excitation-induced\cite{Yoshida2008}
 and exchange-correlation-induced\cite{Yoshida1985}
 negative effective U;
 such as 2Ag$^{2+}$ (d$^9$)
 $\rightarrow$ Ag$^+$ (d$^{10}$) + Ag$^{3+}$ (d$^8$)
 and 2Au$^{2+}$ (d$^9$)
 $\rightarrow$ Au$^+$ (d$^{10}$) + Au$^{3+}$ (d$^8$)\cite{Son2005}
 upon the hole-doping.

\section*{Acknowledgement}
The authors acknowledge the financial support from
 the Global Center of Excellence (COE) program "Core Research and
 Engineering of Advanced Materials - Interdisciplinary Education Center for
 Materials Science", the Ministry of Education, Culture, Sports, Science and
 Technology, Japan, and a Grant-in-Aid for Scientific Research on Innovative
 Areas "Materials Design through Computics: Correlation and Non-Equilibrium Dynamics".
We also thank to the financial support from the Advanced Low Carbon Technology Research and 
Development Program, the Japan Science and Technology Agency for the financial support.




\bibliographystyle{elsarticle-num}
\bibliography{bibfile}







\end{document}